\documentclass[fleqn,usenatbib]{mnras}

\usepackage{newtxtext,newtxmath}
\usepackage[dvipdfmx]{xcolor}
\usepackage{autobreak}
\usepackage{upgreek}

\usepackage[T1]{fontenc}

\DeclareRobustCommand{\VAN}[3]{#2}
\let\VANthebibliography\thebibliography
\def\thebibliography{\DeclareRobustCommand{\VAN}[3]{##3}\VANthebibliography}

\usepackage{graphicx}	
\usepackage{amsmath}	

\newcommand{\beq}{\begin{equation}}
\newcommand{\eeq}{\end{equation}}
\newcommand{\bespl}{\begin{split}}
\newcommand{\enspl}{\end{split}}

\title[Rapidly rotating substellar objects]{A numerical modeling of rotating substellar objects up to mass-shedding limits}
\author[S. Yoshida]{
Shin'ichirou Yoshida\thanks{E-mail: yoshida@ea.c.u-tokyo.ac.jp}
\\
Department of Earth Science and Astronomy, Graduate School of Arts and Sciences, 
The University of Tokyo, 3-8-1 Komaba, Meguro-ku, Tokyo 153-8902 Japan
}

\date{Accepted XXX. Received YYY; in original form ZZZ}

\pubyear{20xx}

\begin{document}
\label{firstpage}
\pagerange{\pageref{firstpage}--\pageref{lastpage}}
\maketitle

\begin{abstract}
Rotation may affect the occurrence of sustainable hydrogen burning in very low-mass stellar objects by the introduction of centrifugal force to the hydrostatic balance as well as by the appearance of rotational break-up
of the objects (mass-shedding limit) for rapidly rotating cases. 
We numerically construct the models of rotating very low-mass stellar objects that may or may not
experience sustained nuclear reaction (hydrogen-burning) as their energy source.
The rotation is not limited to being slow so the effect of the rotational deformation
of them is not infinitesimally small. 
Critical curves of sustainable hydrogen burning in the parameter space of mass versus central degeneracy, on which the nuclear energy generation balances the surface luminosity, are obtained for different values of angular momentum.
It is shown that if the angular momentum exceeds the threshold $J_0=8.85\times 10^{48}{\rm erg}~{\rm s}$
the critical curve is broken up into two branches with lower and higher degeneracy because of the mass-shedding limit.
Based on the results, we model mechano-thermal evolutions of 
substellar objects, in which cooling, as well as mass/angular momentum
reductions, are followed for two simplified cases. The case with such external braking mechanisms as magnetized wind or magnetic braking is mainly controlled
by the spin-down timescale. The other case with no external braking leads to
the mass-shedding limit after gravitational contraction. Thereafter the object
sheds its mass to form a ring or a disc surrounding it and shrinks.

\end{abstract}




\section{Introduction}

Brown dwarfs are substellar mass objects that are too light to have sustainable hydrogen
burning \citep{1962AJ.....67S.579K, 10.1143/PTP.30.460}.
These substellar objects are thought to cool and contract under
its gravity in Kelvin-Helmholtz (K-H) timescale, in which gravitational
binding energy is radiated away. 
If the angular momentum of the object is conserved during its contraction, its rotational frequency increases (it spins up). Even if the angular momentum is lost from an object it may still spin up if the loss is compensated by the decrease in moment of inertia.
Thus it is expected that
some brown dwarfs may be rotating close to a limit beyond which the stellar
matter at the equatorial surface is no longer bound to the object. This
limit is called the mass-shedding limit. 
A dwarf star with an ultra-short period is reported in \cite{2016ApJ...821L..21R}
where radio flares are observed to be periodic in 0.288 hours. This period
is, however, controversial and other authors report a longer period
for the same objects \citep{2017ApJ...834..117W}.
The shortest spin 
period of brown dwarfs is currently one hour, which amounts to several
tens of percent of the angular frequency of mass-shedding limit \citep{2021AJ....161..224T}. 
Although the sample of the three dwarfs
reported in the paper with roughly the same spin period is remarkable and may suggest some unknown mechanism to
suppress further spin up, it is not at all conclusive (see e.g., \citealt{2014prpl.conf..433B}). It is thus relevant to study
the hydrostatic equilibrium figures of rotating substellar objects up to the mass-shedding limit.

Modeling of non-rotating substellar objects has a long history
for sixty years and very elaborate models exist 
\citep{1996A&A...308L..29T, 1997A&A...327.1039C,1998A&A...337..403B, 2012RSPTA.370.2765A, 2015A&A...577A..42B} that take into account
1) detailed equations of state (EOS) of partly degenerate interior
and molecular components in the outer layer; 2) opacity of molecular hydrogen,
metals and dust components; 3) surface convective energy flux;
4) cloud formation in the atmosphere.
On the other hand simplified semi-analytic models of the substellar object \citep{1993RvMP...65..301B, 2001RvMP...73..719B, 2016AdAst2016E..13A, 2019ApJ...871..227F} are yet quite useful in elucidating qualitative properties
of these low-mass objects. In fact \cite{2019ApJ...871..227F} discuss 
possibility of creating brown dwarfs more massive than the minimum mass
of the critical curve in the mass-central degeneracy space, on which the nuclear energy generation of an object is balanced by the thermal radiation from the surface
 (see also the preceding ideas by \citealt{1992ApJ...393..258S, 1999ApJ...517L..39H, 2001ApJ...558....1L}). 
%
According to them, the evolution of some (possibly rare) 
		close binaries composed of 
		brown dwarfs due to gravitational radiation may result in a stable 
		accretion from the secondary through Roche lobe overflow. 
The primary which is initially lighter than 
the minimum mass for stable hydrogen burning may cool down, 
contract, and gain weight without fusing the hydrogen if the mass accretion occurs sufficiently slowly compared with the cooling timescale. Eventually, the object
may still be below the critical curve (thus staying as a brown dwarf), but with its mass larger than the
minimum of the critical curve.
\footnote{In a more recent study by the same group, it is argued an
over-massive brown dwarf may more likely form 
in an evolutionary path of close binaries of AGB-brown dwarf
driven by the AGB wind accretion \citep{2022ApJ...932...91M}.}

The effect of rotation on modeling substellar stars has not been investigated 
thoroughly, although there may exist rapidly rotating objects for which the rotational effects may not be negligible. \cite{2022ApJ...929..117C} may
be a recent exception that studies the minimum mass models of rotating
objects that reach stable hydrogen burning at the center. They found a
fitting formula for the minimum mass of hydrogen burning as a function of
angular frequency. 

They argue that because of the mass-shedding effect, there appears a 
maximum mass of the low mass object that eventually starts hydrogen
burning. A star with a larger mass than this maximum directly starts
its life as a low-mass main sequence, without experiencing the prelude
as a brown dwarf.


In this study, we work on the numerical construction of equilibrium figures of substellar and very low-mass stars that takes into account non-perturbative
effect of rotation. We do not rely on a polytropic approximation of the bulk
of star as in \cite{2016AdAst2016E..13A, 2019ApJ...871..227F,
2022ApJ...929..117C}, but we construct EOS of finite entropy gas
composed of electrons, ions, neutral atoms, and photons (see Sec.\ref{sec: EOS interior}). This is because the higher-order finite entropy effect neglected in those
preceding studies may result in a significant deviation of EOS in the hottest
core region near the threshold of hydrogen burning.
	Here we compute the ratio of the nuclear energy generation rate
	to the surface luminosity of our models. The ratio defines
	critical curves in the parameter space of the degeneracy
	and the mass, on which the ratio is unity. These curves
	are generalizations of the similar curves in \cite{2016AdAst2016E..13A} and \cite{2019ApJ...871..227F} to the rotating objects.
	Above these curves in the parameter space, a model may
	adjust its thermal structure to evolve into a very low-mass main sequence star. Below these curves, the energy loss by the surface
	radiation is not compensated by the nuclear energy generation and
	the object contracts in its Kelvin-Helmholtz timescale.
 By using these critical curves
we also present simple evolutionary models of brown dwarfs.

\section{Formulation}\label{sec: formulation}
\subsection{Assumptions}
To construct the equilibrium models of brown dwarfs, we make the following assumptions.
First, the models are stationary since the timescale of the thermal evolution of the stars is much longer than the hydrodynamical timescale. Also, we are interested in the single
star model in rotation, thus the models are axisymmetric around their rotational axis.
Second, the stars develop full convection in their interior since the mass of the models
are much smaller than that of the Sun. Consequently, the stars are assumed to rotate uniformly.
Also, specific entropy inside the star is constant.
Third, the bulk of the star contains partially degenerate electrons
which contribute mainly to the gas pressure, although we take into
account other pressure contributions (see below). 
The correction to pressure from Coulomb interaction scales as $n_e^{4/3}$
	  (\citealt{1983bhwd.book.....S, 2007coaw.book.....C}), where $n_e$ is the number
	  density of electrons. In the stellar interior, the ratio between degenerate pressure
	  contribution from free electron ($\sim n_e^{5/3}$) to the Coulombic correction scales as $n_e^{1/3}$. Thus the correction is
	  negligible in the interior. On the other hand near the surface of the star, the pressure
	  is mainly from the ideal gas component whose pressure is proportional to $Tn_e$, where
	  $T$ is the temperature of the gas. The ratio of the Coulomb correction to pressure
	  to that of the ideal gas thus scales as $n_e^{1/3} T^{-1}$. Thus it becomes negligible
	  for smaller $n_e$ at finite $T$. Therefore we neglect the Coulomb correction
	  in this study. 
Finally, we neglect the magnetic field for
simplicity.

\subsection{Hydrostatic equilibrium}
Our model objects consist of a bulk part which is mainly supported by
the partially degenerate electron pressure and of a geometrically thin surface photosphere. The radial extent of the latter is so small that it is treated as
a boundary of the bulk interior.
From the assumption above, we may compute the equilibrium models of the bulk of the brown
dwarf by using Hachisu's self-consistent field (HSCF) method \citep{1986ApJS...61..479H}. 
With the assumption above, we can cast the equations of hydrostatic
equations of a stationary and axisymmetric object with Newtonian self-gravity as follows. The equations to be solved are the first integral of 
the momentum equation,
	\beq
		\int\frac{dp}{\rho} + \Phi - \int\Omega^2 RdR = C,
	\eeq
	where $p, \rho$ is the pressure and the density, $\Phi$ is the
	 gravitational potential, $\Omega$ is the angular frequency
	 of the star, $R$ is the cylindrical distance from the rotational axis.
	 $C$ is the integration constant. The gravitational potential is
	 the solution of the Poisson's equation, which is treated in its
	 integrated form with the Green's function,
	 \beq
	 	\Phi({\bf r}) = -G\int\frac{\rho({\bf r}')}{|{\bf r}-{\bf r'}|} d{\bf r'}.
	 \eeq
The equations are solved on finite grid points in the two-dimensional
	 	domain of the upper half of the meridional section.
For a given equation of state, we may iteratively solve these equations
for the stellar models from non-rotating to the mass-shedding limit. The stellar rotation
is parametrised by the ratio ${\rm ax}\equiv R_p/R_e$ of polar surface radius $R_p$ to the equatorial surface
radius $R_e$. 
A better numerical convergence is possible in HSCF compared with other
self-consistent-field method to compute equilibria of rotating stars,
with the axis ratio ${\rm ax}$ of the object being fixed as a constraint.
To compute the stellar interior model, we have modified our
numerical code used in \cite{2019MNRAS.486.2982Y, Yoshida2021}.

For the stellar interior, we assume the simple model of a fully ionized mixture of
hydrogen and helium, while we assume the photosphere is composed of
hydrogen molecule ${\rm H_2}$, neutral and ionized hydrogen,
electrons, and neutral helium. This is because the typical surface temperature of low-mass stars is lower than the ionization energy of helium.

Our assumption of a fully convective model set the constraint that the specific entropy
of the matter inside the star and the photosphere have the same value.

The stellar photosphere is very small in mass compared with
the stellar interior and its thickness is very small compared with the radius.
Therefore we solve for the stellar interior assuming the photosphere has a negligible effect on it, since we only need to estimate the photospheric temperature to evaluate the luminosity. More precisely, the stellar interior is computed by imposing zero enthalpy boundary condition. 
The optical depth in the atmosphere is computed
		by integrating the opacity from the surface inward. 
		We here assume a grey atmosphere.
		Then we determine the photospheric points and 
		the thermodynamic variable there, by equating the
		optical depth to be $2/3$ \citep{2016AdAst2016E..13A, 
		2019ApJ...871..227F}. See Sec.\ref{sec:luminosity}
		for detail.


\subsection{Equation of state}
\subsubsection{Stellar interior}\label{sec: EOS interior}
The stellar matter is assumed to be composed of ionized hydrogen, helium, and metals as well as 
partially degenerate electrons. Electrons are partially degenerate in this
mass range of objects. Therefore the number density $n_e$, the pressure $p_e$,
and the internal energy density $u_e$ of electrons are
expressed by Fermi-Dirac integrals $F_k$ as \citep{1968pss..book.....C},
	\beq
		n_e = \frac{8\uppi\sqrt{2}}{h^3}m_e^3c^3\theta^{3/2}
		[F_{1/2}(\eta, \theta)+\theta F_{3/2}(\eta, \theta)]
	\eeq
	\beq
		p_e = \frac{16\mathrm{\upi}\sqrt{2}}{3h^3}m_e^4c^5\theta^{5/2}
		[F_{3/2}(\eta, \theta)+\theta/2 F_{5/2}(\eta, \theta)]
	\eeq
	\beq
		u_e = \frac{8\mathrm{\upi}\sqrt{2}}{h^3}m_e^4c^5\theta^{5/2}
		[F_{3/2}(\eta, \theta)+\theta F_{5/2}(\eta, \theta)]
	\eeq
	where 
	\beq
		F_k(\eta, \theta) =\int_0^\infty \frac{x^k(1+x\theta/2)^{1/2}}{e^{x-\eta}+1}dx.
	\eeq
Here $h$ is the Planck constant, $\theta = k_BT/m_ec^2$ is the dimensionless temperature and $\eta = \mu/k_BT$ is the degeneracy parameter, where $k_B$ is the Boltzmann constant,
$m_e$ is the electron mass, $c$ is the speed of light, $T$ is the temperature of
electrons and $\mu$ is the chemical potential of electrons. 
In the presentation of the results we use $\psi\equiv\eta^{-1}$ as a parameterization
of degeneracy following the preceding studies \citep{2016AdAst2016E..13A, 2019ApJ...871..227F,
2022ApJ...929..117C}. Thus the larger $\psi$ means the higher temperature and the weaker degree of degeneracy. 
We fix the central density and temperature to determine such other thermodynamic parameters as specific entropy, chemical potential, and degeneracy parameter $\eta$ and $\Psi$. As for other parameters to specify an equilibrium model, we choose the axis ratio ${\rm ax}$.
In the preceding works above, thermodynamic variables of electrons are 
obtained by expanding the Fermi-Dirac integrals in a series of $\psi$. The
expansion is truncated at the linear level in \cite{2022ApJ...929..117C},
while they retain the quadratic terms in \cite{2016AdAst2016E..13A}
and \cite{2019ApJ...871..227F}. If we assume a typical 
central density of very low-mass stars as $\sim 100{\rm g}{\rm cm}^{-3}$
and central temperature as $\sim 3\times 10^6$(K), 
\footnote{
We took these parameters from Fig.5 in \cite{1985ApJ...296..502D}.
}
we have $\psi^2\sim 0.5$. 
The only preceding study dealing with rotating equilibria
		\citep{2022ApJ...929..117C} may be affected by this
		truncation error of EOS.
%
%
We, however, do not follow
this procedure and compute the EOS by assembling the contribution of each species
and by evaluating the Fermi-Dirac integrals numerically. 
l
For the numerical evaluation of Fermi-Dirac integrals,
 we use the Fortran routine available from Frank Timmes' {\tt Cococubed} \citep{1999ApJS..125..277T}.
 \footnote{{\tt https://cococubed.com/code\_pages/fermi\_dirac.shtml}}

Specific entropy of electrons is then given as
	\beq
		s_e = \frac{u_e+p_e}{\rho T} - \frac{\eta n_e k_B}{\rho}
	\eeq
where mass density $\rho$ is
	\beq
		\rho = \biggl\langle\frac{A}{Z}\biggr\rangle m_{_H} n_e
	\eeq 
	where $m_{_H}$ is the atomic mass unit and $\langle A/Z \rangle$ is the
	averaged mass of baryon per electron. 
	In this study, we fix the composition of gas as hydrogen mass fraction $X=0.711$ 
	and helium mass fraction $Y=0.2741$ \citep{2003ApJ...591.1220L}.
	The metal component is so small that we neglect below to compute 
	thermodynamic quantities of the stellar interior.
	For the gas with the mass fraction of hydrogen $X$
	and the helium $Y$, $\langle A/Z\rangle = (X + Y/2)^{-1}$.
	
As for the ions we adopt the equation of state of an ideal gas, thus the pressure
$p_I$ is
	\beq
		p_{_I} = \frac{\rho}{m_H}\left(X + \frac{Y}{4}\right) k_B T.
	\eeq
	The entropy of ions is computed with the Sackur-Tetrode formula
	assuming monatomic molecules \citep{greiner2012thermodynamics},
	\beq
		s_I / k_B = X\left(\frac{5}{2} + \ln W_{\rm H}\right) + \frac{Y}{4}\left(\frac{5}{2}+\ln W_{{\rm H_e}}\right).
	\eeq
Here, $W_{\rm H}$ and $W_{{\rm H_e}}$
		are defined as,
	\beq
		W_i = \frac{(2\mathrm{\upi} m_i k_BT)^{3/2}}{h^3 n_i},
	\eeq
where $m_i (i={\rm H}, {\rm H_e})$ is the mass of ions and $n_i$ is 
the number density of ions. 

We also add the contribution to pressure $p_\gamma$ 
and entropy $s_\gamma$ from photons expressed as
	\beq
		p_\gamma = \frac{a}{3}T^4,
	\eeq
and
	\beq
		s_\gamma = \frac{4a}{3\rho}T^3,
	\eeq
where $a$ is the radiation constant.

Because the star develops full convection inside, we constrain the thermodynamic
variables by fixing total specific entropy to obtain the equation of state. Our numerical procedure is that we iteratively solve for density, pressure, temperature, 
and chemical potential as functions of enthalpy on its finite grid points,
\footnote{This is because the primary variable to be solved in the HSCF scheme
is enthalpy. See \cite{1986ApJS...61..479H}.}
then the cubic interpolations of the thermodynamic variables are performed to create
tables of the equation of state.

\subsubsection{Equation of state near the stellar surface}
Since the low mass stellar models considered here have low surface density and temperature, we assume the gas is composed of partially ionized hydrogen, electron, and helium as well as molecular hydrogen (${\rm H}_2$). Neutral
hydrogen and helium are assumed to form monatomic gas and the Sackur-Tetrode formula per particle,
	\beq
		s/k_B = \frac{5}{2} + \ln\left(\frac{(2\mathrm{\upi} mk_BT)^{3/2}}{h^3 n}\right)
	\eeq
is adopted to compute entropy, 
where $m$ and $n$ are the mass and number density of the particle.

For hydrogen molecules, we also add the contribution $s_{\rm rot}$ from the rotational degree of freedom of the molecule
\citep{greiner2012thermodynamics},
	\beq
		s_{\rm rot}/k_B = 1 + \ln\left(\frac{4\mathrm{\upi}^2 I_1k_B T}{h^2}\right),
	\eeq
where $I_1 = md^2/2$ is the moment of inertia of the hydrogen molecule
and $d=3.7\times 10^{-9}$cm is the approximate length of the molecular bond.
For the typical surface temperature of the low mass objects, vibrational degrees
of freedom of ${\rm H}_2$ molecules are frozen.

The ionization fraction $x_{{\rm H}^+} = n_{{\rm H}^+}/n_{\rm H}$
is obtained by solving the Saha-Boltzmann relation,
	\beq
		\frac{n_{{\rm H}^+} n_e}{n_{\rm H}} = \frac{(2\mathrm{\upi} m_e k_B T)^{3/2}}{h^3}
		\exp\left({-\frac{I_H}{k_B T}}\right),
	\eeq
where $I_H=2.18\times 10^{-11}$ergs is the ionization energy of hydrogen.

The molecular fraction of hydrogen $x_{{\rm H}_2}=n_{{\rm H}_2}/n_{\rm H}$ is obtained by solving the Saha-Boltzmann equation
	\beq
		\frac{n_{\rm H}^2}{n_{{\rm H}_2}} = \frac{(\mathrm{\upi} m_{_H} k_B T)^{3/2}}{h^3}
		\exp\left(-\frac{D}{k_B T}\right),
	\eeq
where $D=4.47$eV is the dissociation energy of ${\rm H}_2$.

We match the specific entropy at the surface with the one inside the star.
In \cite{2016AdAst2016E..13A} (who follow \citealt{1992ApJ...391..817C}) 
it is argued that the first-order phase transition of hydrogen may occur
near the surface, which introduces discontinuity of entropy there.
We here simply neglect this possibility and match the specific
entropy inside and at the surface (see also \citealt{2019ApJ...871..227F}).

By fixing the total entropy value, we construct the adiabatic equation of state 
assuming the ideal gas equation of state,
	\beq
		p = \frac{\rho k_B T_s}{\mu_s m_{_H}},
		\label{eq:surface-adiabatic-EOS}
	\eeq
where $T_s$ is temperature and $\mu_s$ is the mean molecular weight,
	\beq
		\mu_s^{-1} = (1-x_{{\rm H}_2}) X (1+x_{{\rm H}^+}) + x_{{\rm H}_2}X/2 + Y/4.
	\eeq 
\subsection{Onset of sustainable hydrogen burning}
In this study, we consider equilibrium models 
		in the parameter space
		of the degeneracy $\psi$ and the mass $M$.
		When another parameter that characterizes the 
		rotation (i.e., angular momentum) is also chosen, 
		we can specify a single model.
To determine if a model object becomes a main sequence star with
a sustainable nuclear burning, we need to compute the nuclear energy
generation rate $L_n$ and the surface luminosity $L_s$. The ratio of
two, $L_n/L_s$, determines the nature of the objects. 
The critical curve is defined as a curve on 
		which $L_n/L_s=1$ is satisfied in the parameter space.
When the equality $L_n/L_s=1$ is established, the energy
radiated away from the surface is compensated by the nuclear 
energy production inside
the object. 
Thus the object is a main sequence star.
If the ratio is larger
than unity, the energy generation inside the star exceeds what can be
radiated from the surface. The object then adjusts its central degeneracy and
the surface temperature so that the energy generation and radiation balance,
that is, the object evolves toward the critical curve.
If the ratio
is smaller than unity, the nuclear fusion cannot halt the gravitational contraction
and the object contracts and
increases its degeneracy.
 
\subsubsection{Surface luminosity}\label{sec:luminosity}
We follow \cite{2016AdAst2016E..13A} (see also \citealt{2019ApJ...871..227F} and \citealt{2022ApJ...929..117C}) for computing the surface luminosity. First we
compute photospheric temperature $T_s$ as follows.
Near the surface, we have the hydrostatic force balance
	\beq
		\frac{dp}{\rho} + d\Phi - d\chi = 0,
	\eeq
where $\Phi$ is the gravitational potential and $\chi = \int \Omega^2 R dR$
with $R$ being the distance from the rotational axis.
Then the optical depth of the stellar atmosphere $\tau$ is estimated as
	\beq
		\tau = \int_r^\infty \kappa_{_R}\rho dr 
		= \frac{\kappa_{_R} p}{\frac{\partial}{\partial r}(\Phi - \chi)},
	\eeq
where $\kappa_{_R}$ is the Rosseland mean opacity of the stellar
atmosphere. We define the photosphere as $\tau = 2/3$, thus
	\beq
		p = \frac{2}{3\kappa_{_R}} \frac{\partial}{\partial r}(\Phi-\chi)
	\eeq
at the photosphere.  By equating the expression of the pressure
above with Eq.(\ref{eq:surface-adiabatic-EOS}), we may iteratively determine $T_s$. 
Radial derivatives of $\Phi$ and $\chi$
at the stellar surface are determined by the HSCF method. As for the Rosseland
mean opacity of low-temperature gas, we interpolate
Table 1 in \cite{2008ApJS..174..504F}, which gives the mean opacity for
the solar metallicity case. 

Since the object is axisymmetric, $T_s$ is a function of polar 
angle $\theta$. We then obtain the surface luminosity as,
	\beq
		L_s = 4\mathrm{\upi} \int_0^{\mathrm{\upi}/2} R_s(\theta)^2
		\sqrt{1+\left(\frac{d}{d\theta}\ln R_s(\theta)\right)^2}
		\sigma_{_{SB}} T_s(\theta)^4 \sin\theta d\theta
	\eeq
where $R_s(\theta)$ is the surface radius in the direction of $\theta$.
$\sigma_{_{SB}}$ is the Stefan-Boltzmann constant.

\subsubsection{Nuclear burning luminosity}
As for the nuclear burning in low mass objects, we adopt 
the reactions involving proton (p) and deuterium (d)
following \cite{2016AdAst2016E..13A} and \cite{2019ApJ...871..227F}.
As for the pp-reaction (pp-I),
$p+p\to d + e^+ +\nu_e$,
its rate $\epsilon_{pp}$ is
	\beq
		\epsilon_{pp} = 2.5\times 10^6\rho X^2T_6^{-2/3}
		\exp(-33.8 T_6^{-1/3})~{\rm erg}~{\rm g}^{-1}{\rm s}^{-1},
	\eeq
and for the pd-reaction, $p+d\to {}^3{\rm He} + \gamma$,
its rate $\epsilon_{pd}$ is,
	\beq
		\epsilon_{pd} = 1.4\times 10^{24}\rho XX_DT_6^{-2/3}
		\exp(-37.2 T_6^{-1/3})~{\rm erg}~{\rm g}^{-1}{\rm s}^{-1},
	\eeq
where $T_6=T/10^6 K$ and $X_D$ is the equilibrium deuterium mass fraction
computed by \cite{2019ApJ...871..227F},
	\beq
		X_D = 1.79\times 10^{-18} X \exp\left(3.4 T_6^{-1/3}\right)
		\frac{Q_{pd}}{Q_{pp}}
	\eeq
where $Q_{pd}=5.494$MeV and $Q_{pp}=1.18$MeV.

Total nuclear energy generation rate $L_n$ is computed as
the integration inside the object,
	\beq
		L_n = \int dV\rho (\epsilon_{pp} + \epsilon_{pd}).
	\eeq

\subsection{Fitting numerical models}
Our numerical code works in the following way. 
For a given parameter set of central density $\rho_c$ and temperature $T_c$
we compute specific entropy at the center $s_c$ and degeneracy parameter $\eta_c$ or its inverse $\psi_c$. Then the EOS of the stellar interior is fixed. We perform the HSCF iteration
for a given value of axis ratio ${\rm ax}$. 

As in \cite{2019MNRAS.486.2982Y, Yoshida2021}, the spherical polar coordinate
is used in the iteration whose grid numbers are 200 in $r$ and 100 in $\theta$.
Expansion of Green's function by Legendre polynomials is used to invert Poisson's 
equation for gravitational potential. The number of Legendre terms is $25$. 
The converged model is used to compute 
such photospheric parameters as temperature and surface luminosity.

Numerical models are obtained on finite number of grid points in the parameters space of $(\rho_c, T_c, {\rm ax})$. Number of grid points
are $7780$. The parameter ranges they cover are,
$78\le \rho_c ~({\rm g}{\rm cm}^{-3})\le 6200$,
$5\times 10^5 \le T_c ~({\rm K})\le 7.5\times 10^6$,
and $0.63 \le {\rm ax} \le 1$.

Since the parameter space is vast to investigate, we make fitting
formulae for physical variables to be studied here. Instead of 
$(\rho_c, T_c, {\rm ax})$, we use $\log_{10}\psi_c$, $M$, and $f_{\rm obl}=1-{\rm ax}$ or $f_{\rm obl}^2$ as independent parameters. The detail of the fitting
is given in Appendix \ref{sec: fitting}.

\section{Results}\label{sec: Results}
\subsection{Rotational effects on 
	the emergence of very low mass main sequence stars}
\begin{figure}
	\includegraphics[width=\columnwidth]{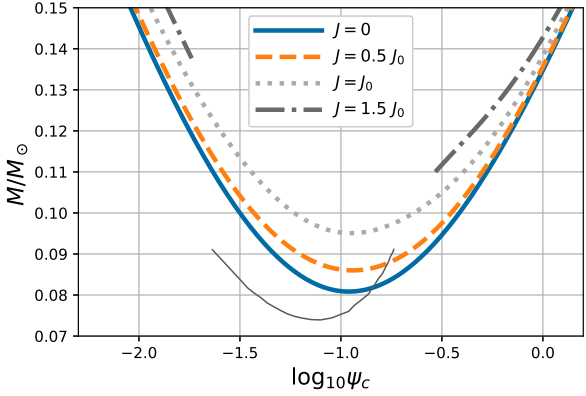}
    \caption{Critical curves of sustained nuclear reactions. Mass at which
    the condition $L_n/L_s=1$ is satisfied is plotted as functions of
    degeneracy $\psi_c$ at the center. On each curve, the angular
    momentum of the models is kept constant. For a model with $J=1.5 J_0$, the bottom
    part of the curve is terminated at the mass-shedding limit, thus the curve
    is split into two branches.
    The gray solid curve is a plot of data taken from Fig.2 of Forbes \& Loeb (2019)
    which assumes the opacity is one-tenth of that of electron scattering.}
    \label{fig: LnLs1}
\end{figure}

One of our main interests is the effect of rapid rotation on
the balance of energy generation by nuclear burning and
radiation from the stellar surface.
We here follow \cite{2019ApJ...871..227F} for the parametrization of the model plot.
%
In Fig.\ref{fig: LnLs1} the critical curves of sustained nuclear reactions
in mass ($M$) versus degeneracy ($\log_{10}\psi_c$) plane 
are plotted for fixed angular momentum. These correspond to Fig.1 of
\cite{2019ApJ...871..227F} for non-rotating cases. 
%
For each value of angular momentum, a model star located above the curve has a nuclear luminosity larger than the surface luminosity. The star may adjust itself to
have higher $\psi_c$ while keeping its total mass constant (moves horizontally
in the diagram), as far as the evolution
proceeds with its mass conserved. Finally, it may sit on one of the critical curves with the same angular momentum. 
For slowly rotating
stars with angular momentum $J\le J_0=8.85\times 10^{48} ({\rm erg}~{\rm s})$ 
the curve is continuous and has a minimum. The minimum corresponds
to the smallest mass of the main sequence star, below which sustainable
hydrogen burning does not take place.

As we increase the angular momentum, the critical curve
is shifted upward and the minimum mass of the main sequence
increases. When the total angular momentum is larger than $J_0$
the critical curve is split into two disjoint segments. This is because
the lost portion of the curve corresponds to the model with
mass-shedding at the equator and cannot be realized as hydrostatic equilibrium.
The leftmost point on the branch with larger $\psi_c$ and the rightmost
point on the branch with smaller $\psi_c$ are the mass-shedding
limit models. 

\begin{figure}
	\includegraphics[width=\columnwidth]{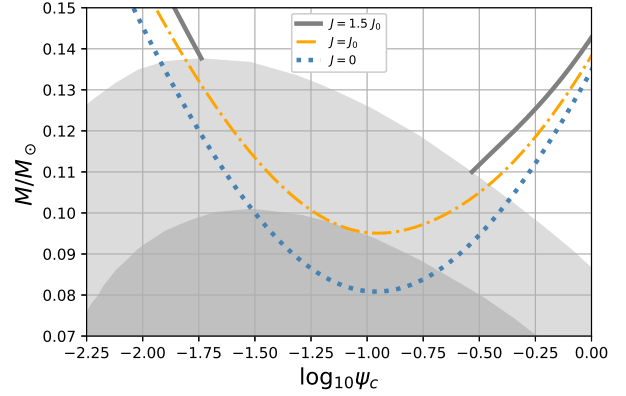}
    \caption{Parameter region of mass-shedding limit for fixed angular momentum
    value and the critical curves. The dash-dotted line is the critical curve
    for $J=J_0$. The dark-shaded region is the parameter region for mass-shedding
    models for $J=J_0$. The light-shaded region is the one for $J=1.5~J_0$.}
    \label{fig: LnLs1-mass-shedding}
\end{figure}

In Fig.\ref{fig: LnLs1-mass-shedding} we show how the
critical curves are split into two branches by the corresponding
parameter region of mass-shedding. The parameter region of
the mass-shedding limit (shaded) expands to a larger mass as angular 
momentum increases. For sufficiently small angular momentum ($J<J_0$)
the critical curve is above this region. The critical curve becomes tangential to
the boundary of the mass-shedding region when $J=J_0$. For larger
angular momentum the critical curve is broken up by the mass-shedding
region.

\begin{figure}
	\includegraphics[width=\columnwidth]{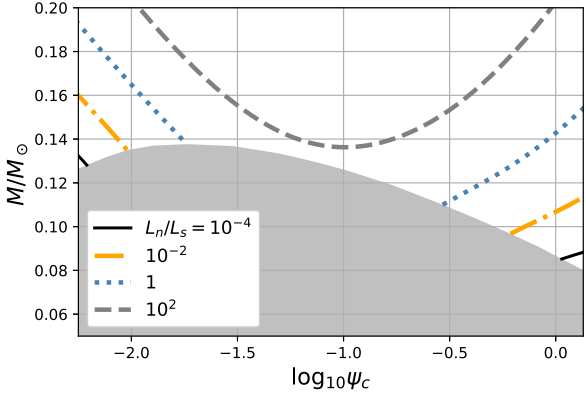}
    \caption{Contour of $L_n/L_s$ value for $J=1.5~J_0$ stars.
    In the hatched region no equilibrium star exists with the fixed value of $J$
    due to mass-shedding}
    \label{fig: LnLsJ_1.5}
\end{figure}

Fig.\ref{fig: LnLsJ_1.5} shows the contour of $L_n/L_s$ ratio
for the fixed angular momentum $J=1.5~J_0$. The mass-shedding region in the parameter space splits the contour lines for 
$L_n/L_s \lesssim 10^2$. The parameter region of a successful hydrogen
burning ($L_n/L_s=1$) is bounded by the mass-shedding region,
therefore the minimum mass for the hydrogen burning is not
determined as an extremum of the critical curve, but as the lighter
model of the mass-shedding limit with $L_n/L_s=1$.
It is therefore expected that a cooling evolution 
of an originally rapidly-rotating brown dwarf would be affected by the mass-shedding limit
(see Sec.\ref{sec: evolutionary paths}).

\begin{figure}
	\includegraphics[width=\columnwidth]{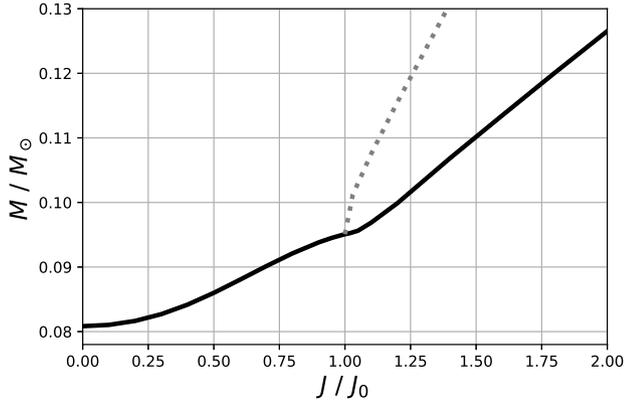}
    \caption{Minimum mass of the critical curve for given angular momentum (solid).
    The dotted line is the minimum mass of the critical curve branch with the stronger 
    degeneracy (e.g., the left branch of $J=1.5J_0$ in Fig.\ref{fig: LnLs1}).
    This branch appears when $J > J_0$.
    }
    \label{fig: MinimumCritical}
\end{figure}

In Fig.\ref{fig: MinimumCritical}, the minimum mass of the critical curve $L_n/L_s=1$
is plotted as a function of the angular momentum. A model below this
curve does not become a main sequence star as far as the angular momentum
is conserved.
We see the mass-shedding limit starts affecting the minimum mass at $J=J_0$, 
where the curve has an inflection point. For $J\ge J_0$,
the mass is limited by the mass-shedding limit.
The dotted line is the mass of the branch with the stronger degeneracy,
which is higher than the one with the weaker degeneracy. Thus the minimum
mass corresponds to the model with the weaker degeneracy and the mass-shedding limit.

\subsection{Simplified evolutionary paths of very low-mass stars and brown dwarfs}\label{sec: evolutionary paths}
Spin-down of very low-mass objects is suggested by 
the comparison of rotational periods of stars in stellar clusters with
different ages (see, e.g., \citealt{2004A&A...421..259S,2005A&A...429.1007S}). The origin of the spin-down mechanism is still debated since the classical spin-down mechanism
of solar-type stars as magnetic breaking or winds may not be so effective in fully-convecting
objects \citep{2014prpl.conf..433B}.
Since the angular momentum loss is not well-understood at present, we here consider two simplified mechano-thermal evolution models of very low-mass objects assuming two extreme cases of angular momentum evolutions.
\subsubsection{Efficient spin-down models}
The first model assumes some external mechanism of spin-down
to be sufficiently effective in the course of the cooling paths. This mechanism
could be either magnetic braking, magnetic stellar wind, or interaction with
a circumstellar disc. We simply parametrize angular momentum loss by introducing
a constant spin-down timescale $\tau_{\rm sd}$. The angular momentum evolution is written as
\beq
	\frac{dJ}{dt} = -\frac{J}{\tau_{\rm sd}}.
	\label{eq: dJdt}
\eeq
The cooling of a star is approximated by 
\beq
	\frac{d\psi_c}{dt} = -\frac{\psi_c}{\tau_{\rm cool}}.
	\label{eq: dUdt}
\eeq
The cooling timescale $\tau_{\rm cool}$ is computed by
\beq
	\tau_{\rm cool}\equiv \frac{U}{L_s},
\eeq
where $U$ is the internal energy of the model. Thus it is a function of
$\psi_c$, $M$, and $J$.
The process takes place in 
Kelvin-Helmholtz (K-H) timescale or spin-down timescale,  
which is anyway much longer than the hydrodynamic timescale. 
The mass loss by winds
may be neglected compared with the angular momentum loss.
We couple and solve Eq.(\ref{eq: dJdt})
and (\ref{eq: dUdt}) assuming total mass is conserved. The surface luminosity $L_s$ is 
computed at each time step. Here we terminate the computation when the stellar
spin is approximately zero (axis ratio becomes 0.995) or it reaches mass-shedding
limit.
\begin{figure}
	\includegraphics[width=\columnwidth]{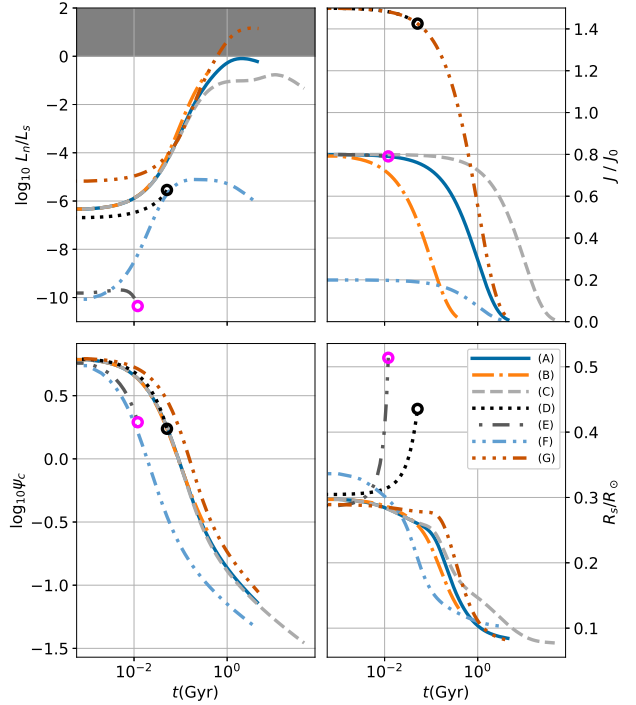}
    \caption{Spin-down evolutions of physical quantities. On each sequence
    the mass in units of $M_\odot$ is fixed while the angular momentum 
    is lost with
    a constant timescale $\tau_{\rm sd}$ (see Eq.(\ref{eq: dJdt}))
    in units of Gyr. The initial
    angular momentum $J_{\rm ini}$ in units of $J_0$ is specified for each curve.
    The parameter set of the curves is the following.
    (A):$(M, J_{\rm ini}, \tau_{\rm sd})=(0.08, 0.8, 1)$ ; (B):$(0.08, 0.8, 0.1)$;
    (C):$(0.08, 0.8, 10)$; (D):$(0.08, 1.5, 1)$; (E):$(0.04, 0.8, 1)$; 
    (F):$(0.04, 0.2, 1)$; (G):$(0.1, 1.5, 1)$. 
    The top left panel shows the luminosity ratio $\log_{10}(L_n/L_s)$.
    The dark-hatched region marks the domain in which an object becomes
    a low-mass main sequence star.
    The top right is the angular momentum plot. The bottom
    left is the degeneracy parameter $\log_{10}\psi_c$. The bottom right
    is the equatorial radius in units of $R_\odot$.
    The open circles at the edge of sequences mark the occurrence of mass-shedding. Otherwise, a sequence terminates at a zero rotation model.}
    \label{fig: spindown-J1}
\end{figure} 
\begin{figure}
	\includegraphics[width=\columnwidth]{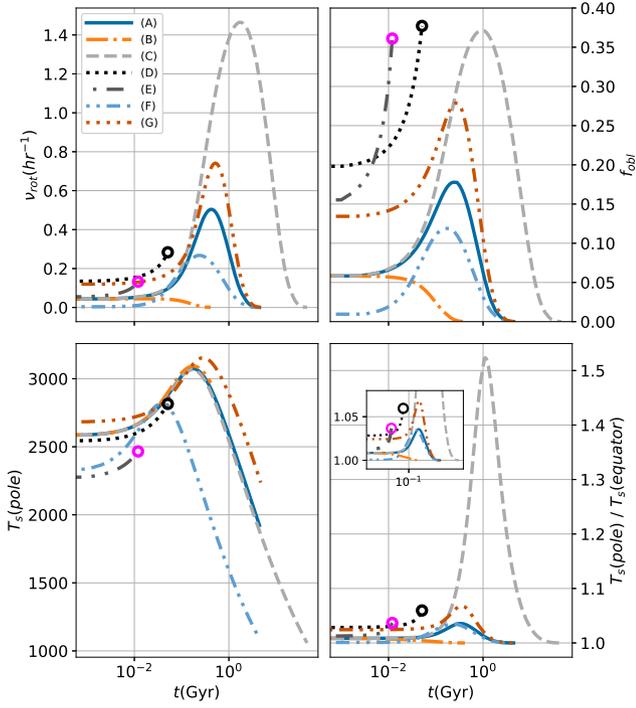}
    \caption{Evolution of additional physical variables for the same models
    as in Fig.\ref{fig: spindown-J1}. The top left panel shows the 
    rotational frequency $\nu_{\rm rot}$ in units of ${\rm hour}^{-1}$.
    The top right is the oblateness. The bottom left is the photospheric
    temperature $T_s$ at the pole. The bottom right is the ratio
    of the polar photospheric temperature to the equatorial one.
    }
    \label{fig: spindown-J2}
\end{figure} 

In Fig.\ref{fig: spindown-J1} and \ref{fig: spindown-J2} spin-down 
evolutions of objects
that conserve masses are exhibited. 
\footnote{Here we split the plots into two groups only
for their visibility.
}
These figures share the same labels (A)-(G) for the models.
On each sequence mass $M$
in units of $M_\odot$ is fixed. The sequences are additionally
parameterized by the initial angular momentum $J_{\rm ini}$ in
units of $J_0$ and the angular momentum loss rate $\tau_{\rm sd}$
in units of Gyr. 
The computation starts at $\log_{10}\psi_c=0.8$. The models
(D) and (E) are initially rotating so rapidly
that they eventually reach the mass-shedding limits which are marked
by the open circles.
The top-left panel of Fig.\ref{fig: spindown-J1} shows the evolution of the logarithmic luminosity ratio. The shaded area corresponds to the stable p-p chain reaction
where an object turns into a main sequence star.
For the low mass models with $M\le 0.08$ it is always negative and
the models never become main sequence stars.
For more massive
cases with $0.1M_\odot$, they reach the nuclear 
burning limit within 1Gyr time for $J_{\rm ini}=1.5J_0$.

As is seen in the bottom-left panel of Fig.\ref{fig: spindown-J1},
the degeneracy monotonically decreases as an object radiates and 
contract. Comparing (A)-(D) and (E)-(F) we see the degeneracy evolution
is mainly determined by the mass, with a very weak dependence
on the initial angular momentum or the spin-down time.
It should be noted that the model (G) reaches the nuclear
burning limit before 1Gyr, after which the degeneracy for these models
would cease to decrease. Our current model, however, does not follow this transition of an object to a very low-mass main sequence star.

The equatorial radius of an object may decrease monotonically
except for the cases that reach the mass-shedding limit.
(bottom-right in Fig.\ref{fig: spindown-J1}).
The radius evolution does not depend strongly on the mass,
on the initial angular momentum, nor on the spin-down time.

In the top-left panel of Fig.\ref{fig: spindown-J2} we show
the evolution of rotational frequency in units of ${\rm hour}^{-1}$.
Rotational frequency changes non-monotonically while the angular
momentum decreases monotonically. This is because the moment
of inertia changes as the structure of the object evolves.
Especially when the spin-down time is large, the star contracts 
while its angular momentum loss is small and it leads to 
a larger increase in the rotational frequency (compare (C) with (A) and (B)).

In the top-right panel of Fig.\ref{fig: spindown-J2}
the oblateness $f_{\rm obl}$ is plotted. The oblateness of low mass
objects may be potentially inferred through observation of polarized
light \citep{2003ApJ...588..545B,2010ApJ...722L.142S}.
For the small $\tau_{\rm sd}$ model  ($\tau_{\rm sd}=0.1$ Gyrs), 
the oblateness monotonically drops down
to zero because of the rapid loss of angular momentum.
It is noticed that the oblateness is not a monotonic function of time
for a case with the larger value of $\tau_{\rm sd}$.
The star initially becomes closer to a sphere but its oblateness
increases and it diminishes again. 
The non-monotonic evolution of $f_{\rm obl}$ may be understood as follows.
We remark that the oblateness scales as $\Omega^2$, where $\Omega$
is the rotational angular frequency, since the stellar deformation is induced
by the centrifugal force. Then we write the total angular momentum $J$
as a function of the stellar radius $R_s$ and $f_{\rm obl}$ as
$J = \alpha M R_s f_{\rm obl}$. The stellar mass $M$ is fixed in the evolution 
and $\alpha$ is a constant factor. 
Then the time derivative of $J$ gives
\beq
	\frac{1}{J}\frac{dJ}{dt} = \frac{1}{R_s}\frac{dR_s}{dt} + \frac{1}{f_{\rm obl}}\frac{df_{\rm obl}}{dt},
\eeq
or
\beq
	\frac{1}{f_{\rm obl}}\frac{df_{\rm obl}}{dt}= -\tau_{\rm sd}^{-1} - \frac{1}{R_s}\frac{dR_s}{dt},
\label{eq: fobl-evolution}
\eeq
since we have $-\tau_{\rm sd}^{-1}$ as a negative constant 
(see Eq.(\ref{eq: dJdt})). When $R_s$ is decreasing
in time (i.e., $dR_s/dt<0$), the sign of the left-hand side is determined by the 
time derivative of $R_s$. We first focus on (A) in Fig.\ref{fig: spindown-J1}
whose spin-down timescale is 1Gyr.
The decline of $R_s$ becomes steeper around $t\sim 0.25$Gyr. Then the
$-d\log R_s/dt$ term becomes large enough to let $f_{\rm obl}$ increases.
After that $f_{\rm obl}$ decreases again as the decline of $R_s$ later slows down.
The model (C) has $\tau_{\rm sd}=10$Gyr. For this model, the first term
on the left-hand side of Eq.(\ref{eq: fobl-evolution}) is smaller than that in (E).
As a result, the duration of increasing $f_{\rm obl}$ is longer.
The model (B) has $\tau_{\rm sd}=0.1$Gyr. This model spins down so quickly
that the term of $-d\log R_s/dt$ has little effect on $f_{\rm obl}$ evolution.
Then $f_{\rm obl}$ decreases monotonically.

The bottom-left panel is the effective temperature at the poles. 
The surface temperature slightly increases at first, then decreases
because the cooling starts to dominate 
the sum of the nuclear energy generation and
the liberation of gravitational binding energy.
The temperature
shows little dependence on $\tau_{\rm sd}$ or on the initial angular momentum. 

As is shown in the bottom-right panel of Fig.\ref{fig: spindown-J2}, the difference between the polar and the equatorial temperature may reach as much as 60\% for the long spin-down time of $10$Gyr, but it is lower than 10\% for the shorter
timescale.

\subsubsection{Decretion models}\label{sec: decretion model paths}
In another limiting case, the spin-down due to magnetic breaking or stellar wind 
may be neglected. Then the star may spin up by contracting while radiating
away its energy,
with the angular momentum being conserved. When the star hits its mass-shedding limit, it sheds 
part of its mass at the surface, which would form a disc/ring around the equator
of the star. This is similar to the so-called 'decretion disc' observed in Be stars \citep{1991MNRAS.250..432L,2001PASJ...53..119O}. The decretion takes place in the K-H timescale
which is much longer than the hydrodynamic timescale. Thus the star sheds its mass
while maintaining hydrostatic equilibrium. 
Here we introduce a simple model of this decretion process. 
	When a star is rotating at the mass-shedding limit, the gas element at the equatorial
	surface is orbiting the star at the Keplerian frequency. Then
	no pressure gradient is necessary for the element to stay there, and the 
	element moves as if it were a test particle in the stellar gravitational field. 
	Once the star cools and contracts, the element is left as it is, since 
	forces acting on it do not change.
	As far as the stellar radius continues decreasing, 
	the gas element orbiting with local Keplerian velocity forms a ring. 
	The evolution of the system is driven by the cooling of the
	object and its timescale is that of K-H contraction. 
	Strictly speaking, the gravitational potential of the whole system may change gradually as the stellar matter is 
	re-distributed to disc. 

We assume a spherical $N=1.5$ polytrope as a model of a uniformly rotating star
neglecting rotational deformation. From the numerical solution of 
Lane-Emden equation for $N=1.5$, we find the mass $M_r$ contained
inside the radius $r$ and the moment of inertia $I_r$ inside the cylindrical
section of radius $r$ scales as $I_r\sim M_r^{\lambda}$ 
with $\lambda\sim 2.5$, for the matter close to the stellar surface.
Suppose a uniformly rotating star with angular frequency $\Omega$
has the mass $M_\star$, the radius $R_\star$, the moment of inertia $I_\star$, and the angular
momentum $J_\star$. When the star reaches the mass-shedding limit,
it loses a small fraction of the outer part whose radial coordinate is
larger than $r ~(<R_\star)$ for some value of $r$. 
Through this decretion process, the amount
of angular momentum lost is $\Delta J = (I_\star - I_r)\Omega$, which 
is simply advected outward with the decreted mass. Then we have
\beq
	\frac{\Delta J}{J_\star} = 1 - \frac{I_r}{I_\star}.
\eeq
Since we have a scaling of $I_r \sim M_r^\lambda$.
\beq
	\frac{\Delta J}{J_\star} = 1 - \left(\frac{M_r}{M_\star}\right)^\lambda
= 1-\left(1-\frac{\Delta M}{M_\star}\right)^\lambda,
\eeq
where $\Delta M$ is the mass decreted. Then for $|\Delta M|/M_\star \ll 1$,
we have
\beq
	\frac{\Delta J}{J_\star} = \lambda \frac{\Delta M}{M_\star}.
	\label{eq: decretion J}
\eeq
 Eq.(\ref{eq: decretion J}) determines the mechano-thermal evolution
 of a star during the period of mass decretion. When the mass decretion
 starts, the star loses its mass and angular momentum. In the $\log_{10}\psi_c - M$ plane, the star then follows a path whose mass and angular momentum
 is determined by Eq.(\ref{eq: decretion J}) and which is at the same time
 sitting on the mass-shedding contour with $J=J_\star - \Delta J$. 
 This decretion process lasts as far as the star stays on a mass-shedding contour 
 for the angular momentum it currently has. The sequence terminates
 either if it is not on the mass-shedding curve or if the radius of the star
 increases. Once the decretion ceases, 
 the star resumes normal K-H 
 contraction by preserving its mass and angular momentum.
 
\begin{figure}
	\includegraphics[width=\columnwidth]{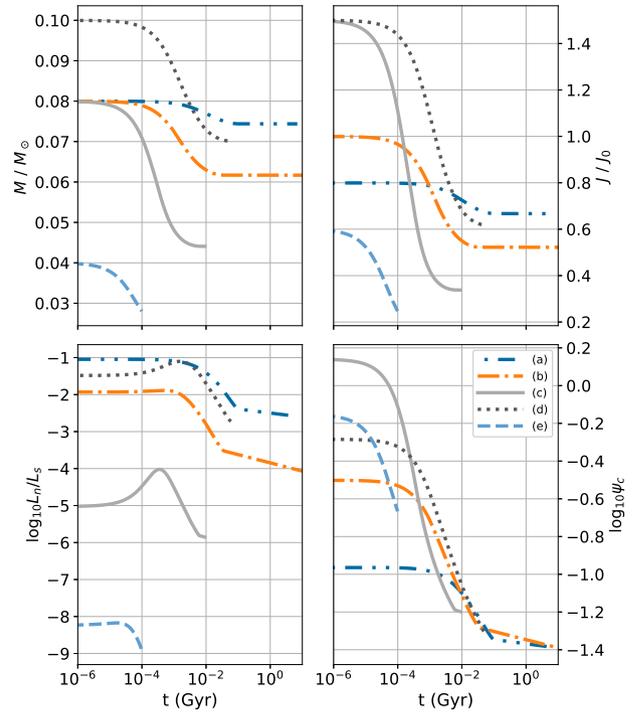}
    \caption{Evolutionary sequences of stars in the decretion processes.
    The parameter set for each curve is the following;
    (a):$(M(t=0), J(t=0))=(0.08, 0.8)$; (b):$(0.08, 1.0)$; (c):$(0.08, 1.5)$;
     (d):$(0.1, 1.5)$; (e):$(0.04, 0.6)$.
    The top left panel shows the mass in units of $M_\odot$. 
    The top right is the angular momentum in units of $J_0$.
    The bottom left
    is the luminosity ratio $\log_{10}(L_n/L_s)$.
    The bottom right is the degeneracy parameter.}
    \label{fig: evolution-decretion1}
\end{figure} 
\begin{figure}
	\includegraphics[width=\columnwidth]{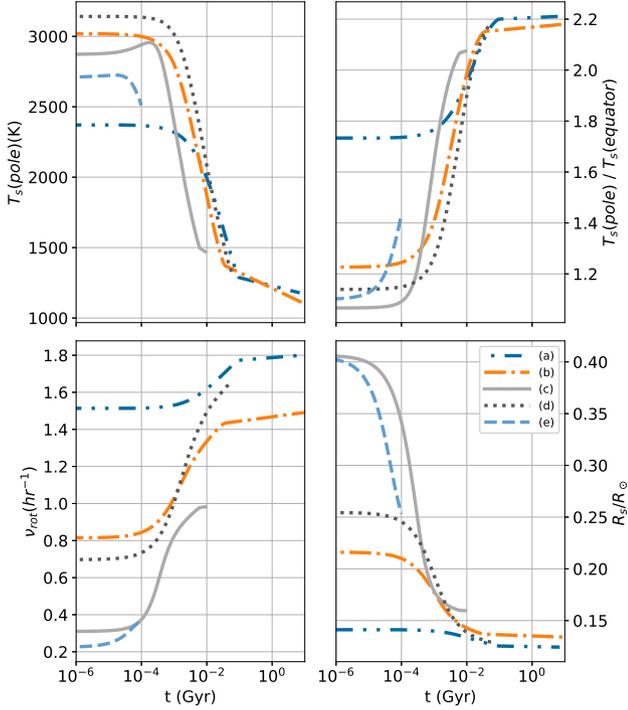}
    \caption{Evolution of additional physical quantities for the same
    models as in Fig.\ref{fig: evolution-decretion1}. The top left
    is the photospheric temperature at the pole. In the top right
    is the ratio of polar to equatorial photospheric temperature.
    The bottom left is the rotational frequency in units of 
    ${\rm hour}^{-1}$. The bottom right is the equatorial radius
    in units of $R_\odot$.}
    \label{fig: evolution-decretion2}
\end{figure} 
In Fig.\ref{fig: evolution-decretion1} and \ref{fig: evolution-decretion2}
evolutions of variables with mass-shedding sequences are shown.
The origin of time is the initiation of the decretion process by mass-shedding.
Each curve terminates at the end of the mass decretion process. 
These models are parametrized by their initial mass $M(t=0)$ and
initial angular momentum $J(t=0)$.

From the top left and right panels in Fig.\ref{fig: evolution-decretion1}
we see a significant fraction of the mass and angular momentum of an original object lost during the whole decretion process. 
Comparing the cases with $M(t=0)=0.08M_\odot$ with different initial
angular momentum, we see that the larger the initial angular momentum,
the stronger the decretion effect is. 
This is because the model with
the larger angular momentum has the larger radius (see the
bottom-right panel of Fig.\ref{fig: evolution-decretion2})
and the larger moment of inertia, which is inferred from
the top-right panel of Fig.\ref{fig: evolution-decretion1}
and the bottom-left panel of Fig.\ref{fig: evolution-decretion2}.
This means the star with the larger $J(t=0)$ has a larger fraction 
of mass and angular momentum in the outer part of it, which
are lost from the star during the decretion process. It means
that for the models with the fixed initial mass, 
the mass and the angular momentum lost in the decretion 
process is larger when the star initially has a larger angular momentum.

As is seen in the bottom-right panel of \ref{fig: evolution-decretion1}
and the top panels of \ref{fig: evolution-decretion2}, the evolution
of the degeneracy and the surface temperature at the later stage
do not depend strongly on the mass nor on the initial angular momentum.

From the top panels of Fig.\ref{fig: evolution-decretion2}, we see
the surface temperature evolution does not strongly depend
on the initial parameters.

A seemingly counter-intuitive trait of the rotational frequency is 
that it is lower for a higher initial angular momentum case, when
the models with the same initial mass are compared.  As is seen
in the bottom-right panel of Fig.\ref{fig: evolution-decretion2},
a star with a larger initial angular momentum has a larger initial radius. 
This is because the centrifugal force makes the star expand
perpendicular to the rotational axis. The moment of inertia becomes
larger for the higher initial angular momentum case and a model
spins more slowly even if it has a larger angular momentum.

Finally, we may speculate on an outcome of the evolution
of the substellar objects through the decretion process.
The result of the decretion process may be a
massive decretion disc surrounding a brown dwarf.
If the disc survives long enough to accumulate a substantial
fraction of the mass of the original dwarf, it may become gravitationally
unstable and fragments into smaller bodies.
The fragmentation may lead to the formation of a binary of brown dwarfs
or a planetary system around a brown dwarf.


\section{Summary and conclusion}
We present numerical models of rotating very low-mass stars and brown dwarfs
up to their mass-shedding limits. Rotation may strongly affect the occurrence of sustainable hydrogen burning in these objects by the introduction of centrifugal force to the hydrostatic balance as well as by the appearance of mass-shedding limit for rapidly rotating cases. Our model takes
into account the non-perturbative effect of rotation on equilibrium figures of
very low-mass objects. 
We obtain the critical curves of sustainable hydrogen burning
in the degeneracy-mass plane, 
on which nuclear reaction rate equals the surface luminosity. 
As is expected, the critical curve for the higher value of angular momentum
shift upward, thus the minimum mass for the stable hydrogen burning
goes up as the angular momentum increases.
There is a limiting value of angular momentum, $J_0=8.85\times 10^{48}$(erg s) beyond which the critical curve is no longer continuous in the parameter plane.
For $J>J_0$, we find the minimum mass for hydrogen burning is constrained by the appearance of the mass-shedding limit.

We also use our models to study mechano-thermal evolutions of substellar
objects that experience the Kelvin-Helmholtz contraction. Two extreme
cases of conservative and non-conservative evolution of mass and angular
momentum are studied. 
For the case of angular momentum loss by some external mechanism such
as stellar wind or magnetic braking, the degeneracy parameter and the surface
temperature do not strongly depend on the parameters except the total mass,
while the stellar rotation frequency and the oblateness depends on the
spin-down timescale. The rotational frequency and the oblateness do not
evolve monotonically. The models having large initial angular momentum
evolve toward the mass-shedding limit.
%

For the case with no external braking mechanism, a rotating model would
follow first a path of Kelvin-Helmholtz contraction with mass and angular momentum being conserved. As it contracts, it may encounter its 
mass-shedding limit. 
If it reaches the mass-shedding limit before it satisfies 
a critical condition for sustained nuclear burning ($L_n/L_s=1$),
the star starts its mass-decretion evolution by losing its excess
angular momentum and mass to the surrounding disc/ring. 
We follow the mass-decretion process from the onset of mass-shedding
limit.
For given initial mass the decretion process takes place
rapidly when the initial angular momentum is larger, which leads to the larger
loss of mass and angular momentum.
It is also remarkable that the rotational frequency monotonically increases
though the angular momentum is lost by the decretion.

The end product of this decretion process may be a small
brown dwarf and a debris disc. Or more interestingly,
if the disc is gravitationally unstable, a binary system of brown dwarfs or
a brown dwarf with a planetary system may form.

\section*{Acknowledgements}
I thank the anonymous reviewer and the scientific editor for their careful reading 
and providing detailed and useful comments that improve the original manuscript.

\section*{Data Availability}
The data underlying this article will be shared upon reasonable request to the corresponding author.


\bibliographystyle{mnras}
\bibliography{bd} 




\appendix
\section{Gravity darkening of rotating very low mass objects}
From von Zeipel's theorem \citep{1924MNRAS..84..665V} the effective surface temperature of a star is a function of surface gravity. Thus in a uniformly rotating
star the surface temperature decreases as we go away from the rotational axis.
The original theorem states that the temperature $T$ and the surface gravitational acceleration $g_{\rm eff}$ which includes the centrifugal contribution
are related as
\beq
T\propto g_{\rm eff}^\beta,
\eeq
where $\beta=0.25$.

\begin{figure}
	\includegraphics[width=\columnwidth]{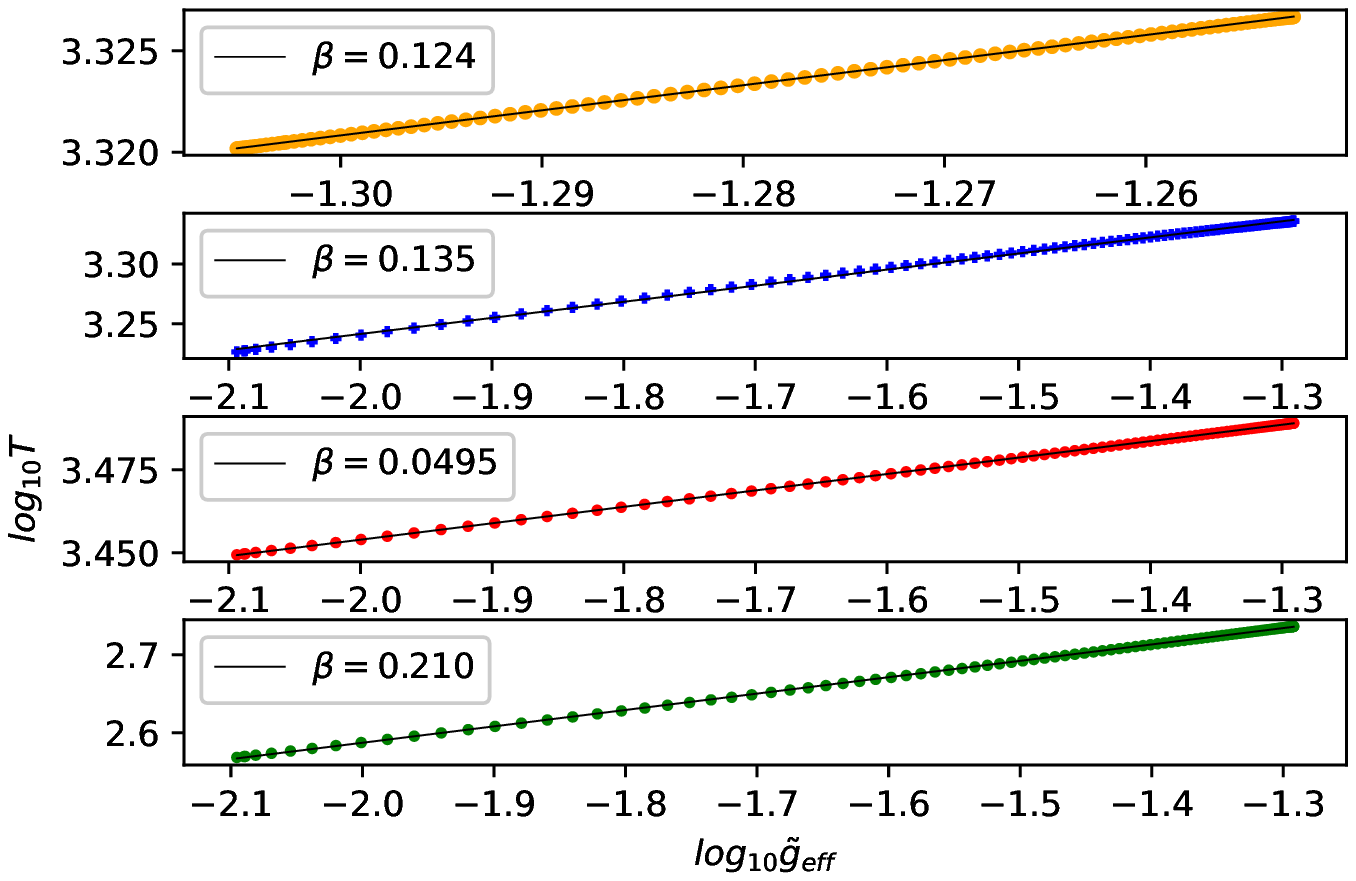}
    \caption{Gravity darkening of our rotating dwarf models. Surface temperature
    in Kelvin is plotted against the local acceleration of effective gravity $\tilde{g}_{\rm eff}$ which is 
    normalized by $4\mathrm{\upi} G\rho_c R_{\rm eq}$, where $\rho_c$ is the central density
    and $R_{\rm eq}$ is the equatorial radius. On each curve, the polar point corresponds to the upper right while the equatorial point is at the lower left.
    The top panel is for a slowly rotating model (axis ratio is 0.96)
    with $\psi_c=0.139$ and $M=0.022M_\odot$. The second panel is for a rapidly rotating model
    (axis ratio is 0.68)
    with the same $\psi_c$ but $M=0.025M_\odot$. The third panel is for a rapidly rotating
    model with $\psi_c=0.788$ and $M=0.082M_\odot$ while the bottom panel is for a
    rapidly rotating model with $\psi_c=0.0185$ and $M=0.073M_\odot$. 
    $\beta$ in each panel is the best-fit power-law index of von Zeipel-like dependence
    of gravity darkening.
    }
    \label{fig: gravity darkening}
\end{figure}
Although the theorem applies only for barotropic stars where pressure depends 
solely on density, it is extended
to the convective case by \cite{1967ZA.....65...89L}. For this case, $T\propto g_{\rm eff}^{0.08}$. By introducing the Roche model and comparing it with their two-dimensional numerical models of rapidly rotating stars, \cite{2011A&A...533A..43E} show that
the gravity darkening is well-represented by a power law neither by von Zeipel's nor Lucy's. Their effective power $\beta$ changes from $\beta\sim 0.25$ 
in non-rotating stars to $\beta \lesssim 0.14$ for the most rapidly rotating stars.

We see the gravity darkening in our rotating models. In Fig. \ref{fig: gravity darkening}
The surface temperature is plotted as a function of local effective gravity $g_{\rm eff}$. 
We see good power-law fittings of $T$ as a function
of $g_{\rm eff}$, though the power index $\beta$ depends on the mass, the rotation, and the degeneracy of the model. For the models with relatively large degeneracy at the center (the first
and second panel in Fig.\ref{fig: gravity darkening}, $\beta\sim 0.12-0.14$
for both slowly and rapidly rotating cases. This suggests that the dependence
of $\beta$ on the axis ratio or the oblateness is rather weak.
Comparing the third and fourth panels, $\beta$ depends strongly on 
the degeneracy parameter.
It should be noted the fitting quality by the power law becomes lower
for rapidly rotating cases as is pointed out by \cite{2011A&A...533A..43E}.
\begin{figure}
	\includegraphics[width=\columnwidth]{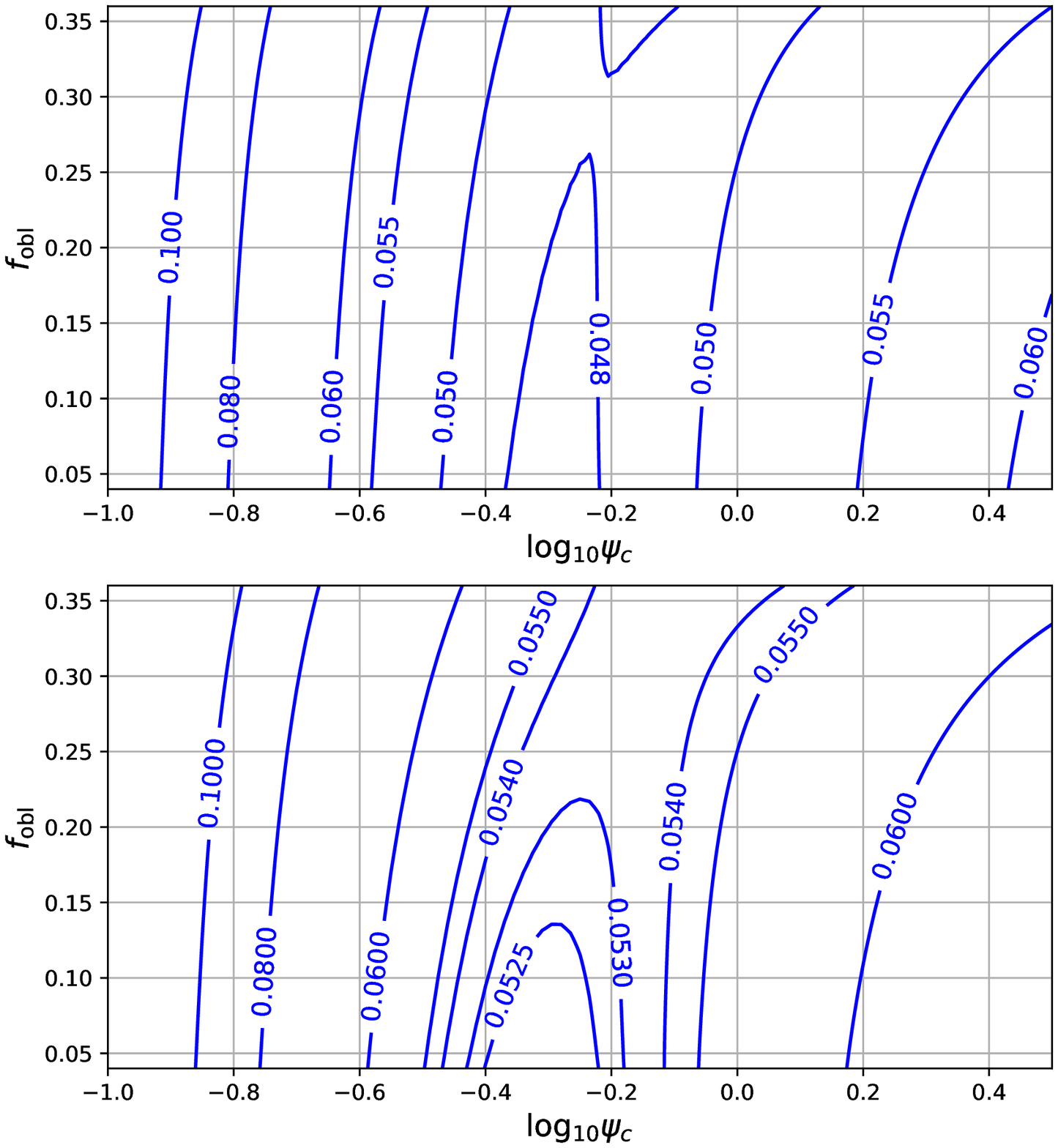}
    \caption{Contour of the power-law index $\beta$ of the gravity darkening law 
    ($T_{\rm eff}\propto g_{\rm eff}^\beta$) in the $\log_{10}\psi_c$ - $f_{\rm obl}$ plane. The top panel is for $M=0.08M_\odot$, while the bottom one
    is for $M=0.04M_\odot$.  
   }
    \label{fig: index beta Mcst}
\end{figure}
\begin{figure}
	\includegraphics[width=\columnwidth]{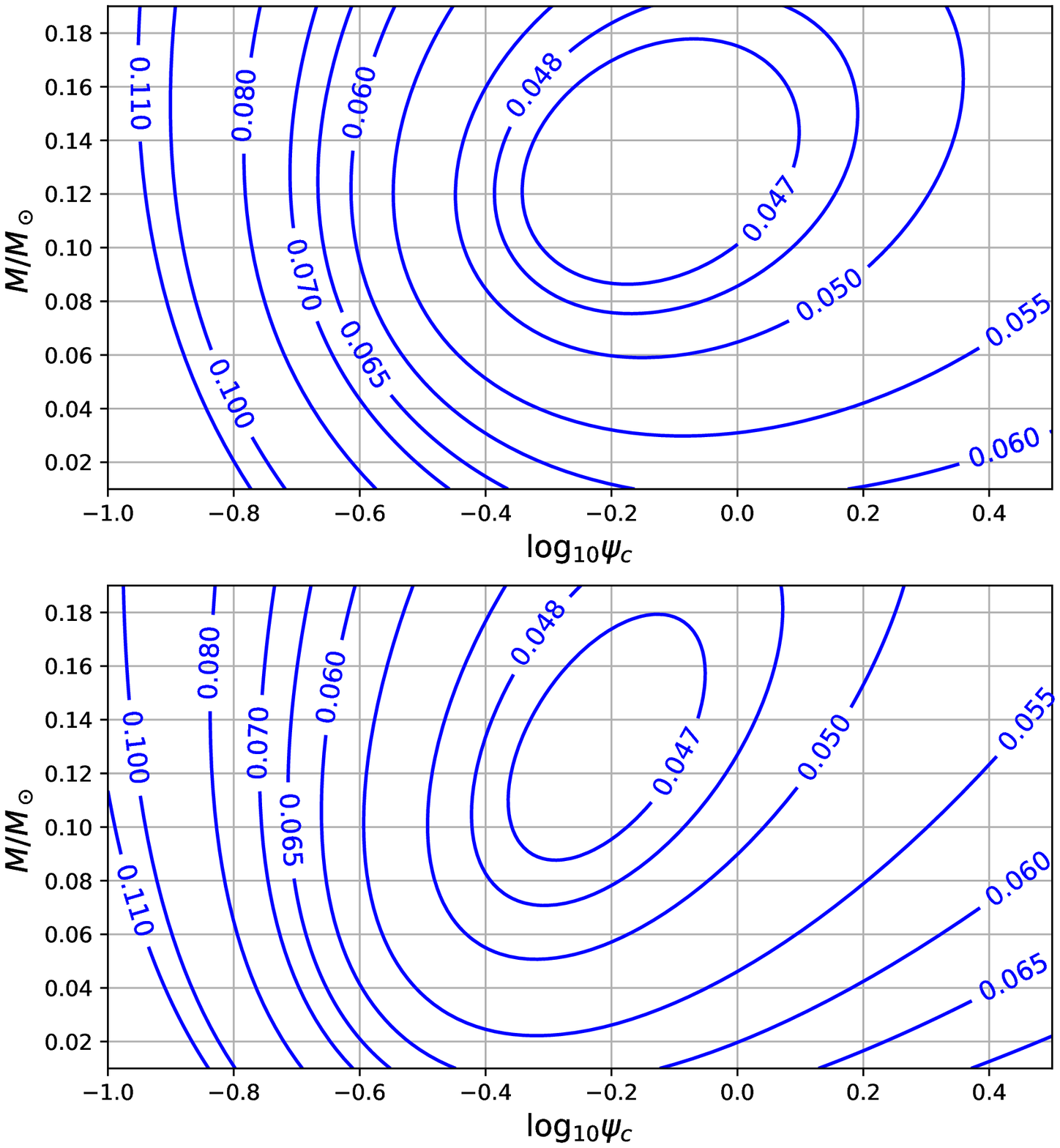}
    \caption{Contour of the power-law index $\beta$  for
    the fixed oblateness $f_{\rm obl}$, The top panel is for $f_{\rm obl}=0.37$
    while the bottom one is for $f_{\rm obl}=0.02$.  
   }
    \label{fig: index beta fcst}
\end{figure}

$\beta$ is in general a function of degeneracy parameter $\psi_c$,
mass $M$, and oblateness $f_{\rm obl}$.
In Fig.\ref{fig: index beta Mcst} the contour of $\beta$ in 
the $\log_{10}\psi_c$ - $f_{\rm obl}$ plane are
given for $M=0.08M_\odot$ (top), and $M=0.04M_\odot$ (bottom).

In Fig.\ref{fig: index beta fcst} the contour in the 
$\log_{10}\psi_c$-$M$ plane are given for $f_{\rm obl}=0.37$
case (top), which are nearly mass-shedding limit,
and for $f_{\rm obl}=0.02$ case (bottom), which are slowly rotating.

\section{Fitting formulae of numerical models}\label{sec: fitting}
The curves in Sec.\ref{sec: Results} are drawn by using fitting formulae of numerical
models. The fitting formulae are defined as functions of $\psi_c$, $M/M_\odot$,
and {\rm ax}, the axis ratio. $\psi_c$ measures how strong the degeneracy of a star is, or how
strong the thermal effect is. $M$ measures the amount of mass and the strength
of gravity. ${\rm ax}$ measures how fast the star is spinning. It should be noted 
that ${\rm ax}$ scales approximately as $\Omega^2$. All the fitting formulae below
reproduce the parameter values computed numerically within the relative error
of 10\%.

Introducing $x=\log_{10}\psi_c$, $y=M/M_\odot$, $z=\sqrt{1-{\rm ax}}$
we fit the angular momentum of the star as, 
\beq
\begin{split}
J/10^{49}({\rm erg~s})&= \left[j_0 + j_1x + j_2y + j_3x^2 + j_4xy + j_5y^2\right]z\\
       & \left[1 + (j_6 + j_7x + j_8y + j_9x^2 + j_{10}xy + j_{11}y^2)z\right]^{-1}
\end{split}
\eeq
\beq
\begin{split}
& \mbox{{\rm where} {\rm fitted} {\rm parameters} {\rm are}}\\
&j_0=0.3360, j_1=0.7962, j_2=16.54, j_3=0.4079, j_4=6.228,\\
& j_5=51.74, j_6=0.2751, j_7=0.1035, \\
&j_8=-4.114, j_9=-0.08969, j_{10}=-1.621,\\
& j_{11}=10.48.
\end{split}
\eeq

$L_n/L_s$ is fitted by using $\zeta=1-{\rm ax}$ instead of $z$ as,
\beq
\begin{split}
\log_{10}\left(\frac{L_n}{L_s}\right) &= 
\left[ l_0(1+b_{01}\zeta+b_{02}\zeta^2)
    + l_1(1+b_{11}\zeta+b_{12}\zeta^2)x\right.\\
&    + l_2(1+b_{21}\zeta+b_{22}\zeta^2)y
    + l_3(1+b_{31}\zeta+b_{32}\zeta^2)x^2\\
&  \left.  + l_4(1+b_{41}\zeta+b_{42}\zeta^2)xy
    + l_5(1+b_{51}\zeta+b_{52}\zeta^2)y^2\right]\\
&    \left[ 1 + l_6(1+b_{61}\zeta+b_{62}\zeta^2)x
    + l_7(1+b_{71}\zeta+b_{72}\zeta^2)y\right.\\
&    + l_8(1+b_{81}\zeta+b_{82}\zeta^2)x^2
    + l_9(1+b_{91}\zeta+b_{92}\zeta^2)xy\\
&   \left. + l_{10}(1+b_{101}\zeta+b_{102}\zeta^2)y^2 \right]^{-1}
         \end{split}
\eeq
\beq
\begin{split}
& \mbox{{\rm where} {\rm fitted} {\rm parameters} {\rm are}}\\
&l_{0}=-19.24 ,l_{1}=-19.90, l_{2}=89.97, l_{3}=-10.19, l_{4}=3.974,\\
 &l_{5}=391.8, l_{6}=1.297,l_{7}=24.57, l_{8}=0.6119, \\
 &l_{9}=4.902,l_{10}=27.84\\
 &b_{01}=-1.306,b_{02}=4.715,b_{11}=-1.235,b_{12}=3.741,b_{21}=-1.699,\\
&b_{22}=6.480,b_{31}=-1.607,b_{32}=4.742,b_{41}=12.07,b_{42}=-39.57,\\
&b_{51}=-2.166,b_{52}=1.313,b_{61}=0.400,b_{62}=-2.504,b_{71}=-1.898,\\
  &b_{72}=4.984,b_{81}=-0.07959,b_{82}=-1.657,b_{91}=-0.9873,b_{92}=1.038,\\
  &b_{101}=-2.176,b_{102}=-1.816.
\end{split}
\eeq

Surface luminosity $L_s/L_\odot$ is fitted by
\beq
\begin{split}
    \log_{10}(L_s/L_\odot) &= 
\left[q_0(1+c_{01}\zeta+c_{02}\zeta^2)
        + q_1(1+c_{11}\zeta+c_{12}\zeta^2)x\right.\\
&        + q_2(1+c_{21}\zeta+c_{22}\zeta^2)y
        + q_3(1+c_{31}\zeta+c_{32}\zeta^2)x^2\\
&        \left.+ q_4(1+c_{41}\zeta+c_{42}\zeta^2)xy
        + q_5(1+c_{51}\zeta+c_{52}\zeta^2)y^2\right]\\
&        \left[1 + q_6(1+c_{61}\zeta+c_{62}\zeta^2)x
          + q_7(1+c_{71}\zeta+c_{72}\zeta^2)y\right.\\
&         + q_8(1+c_{81}\zeta+c_{82}\zeta^2)x^2
          + q_9(1+c_{91}\zeta+c_{92}\zeta^2)xy\\
&   \left.       + q_{10}(1+c_{101}\zeta+c_{102}\zeta^2)y^2 \right]^{-1}
\end{split}
\eeq

\beq
\begin{split}
& \mbox{{\rm where} {\rm fitted} {\rm parameters} {\rm are}}\\
    &q_0=-2.244, q_1=-1.573,  q_2=-17.40, q_3=--0.5792,  q_4=-1.945,\\
    & q_5=-9.714, q_6=0.9424,   q_7=6.946,   q_8=0.2706,\\
               & q_9=2.597, q_{10}=2.012,\\
&c_{01}=0.1424,c_{02}=-0.4922,c_{11}=-0.06819,c_{12}=0.02784,\\
&c_{21}=-1.3035,c_{22}=2.486,c_{31}=1.487,c_{32}=-1.367,\\
&c_{41}=-1.977,c_{42}=0.9328,c_{51}=1.802,c_{52}=-2.004,\\
&c_{61}=-0.3951,c_{62}=0.8406,c_{71}=-0.9667,c_{72}=2.236,\\
&c_{81}=-0.2128,c_{82}=1.000,c_{91}=-0.9093,c_{92}=2.125,\\
&c_{101}=1.444,c_{102}=-1.342.
\end{split}
\eeq

Cooling timescale $\tau_{\rm cool}$ is fitted as
\beq
\begin{split}
  \log_{10}\left(\frac{\tau_{\rm cool}}{\rm 1Gyr}\right) &=  
\left[ t_0(1+d_{01}\zeta+d_{02}\zeta^2)
   + t_1(1+d_{11}\zeta+d_{12}\zeta^2)x\right.\\
   & + t_2(1+d_{21}\zeta+d_{22}\zeta^2)y
    + t_3(1+d_{31}\zeta+d_{32}\zeta^2)x^2\\
   &\left. + t_4(1+d_{41}\zeta+d_{42}\zeta^2)xy
    + t_5(1+d_{51}\zeta+d_{52}\zeta^2)y^2 \right]\\
   & \left[1 + t_6(1+d_{61}\zeta+d_{62}\zeta^2)x
    + t_7(1+d_{71}\zeta+d_{72}\zeta^2)y \right.\\
   & + t_8(1+d_{81}\zeta+d_{82}\zeta^2)x^2
    + t_9(1+d_{91}\zeta+d_{92}\zeta^2)xy\\
   &\left. + t_{10}(1+d_{101}\zeta+d_{102}\zeta^2)y^2 \right]^{-1}.
 \end{split}
\eeq

\beq
\begin{split}
& \mbox{{\rm where} {\rm fitted} {\rm parameters} {\rm are}}\\
&t_0=-2.945, t_1=-2.772, 
    t_2=14.80, t_3=-0.4297, t_4=-0.3028,  t_5=-3.997,\\
    &t_6=0.8193,  t_7=8.680,  t_8=0.2258,
    t_9=2.969,  t_{10}=-2.106\\
    &d_{01}=-0.2074,d_{02}=0.6923,
d_{11}=-2.353,d_{12}=4.824,d_{21}=2.790,\\
&d_{22}=-6.338,d_{31}=-12.00,d_{32}=22.26,
d_{41}=-52.46,d_{42}=126.0,\\
&d_{51}=56.96,d_{52}=-121.9,d_{61}=-2.334,d_{62}=4.386,
d_{71}=-2.896,\\
&d_{72}=5.579,d_{81}=-2.244,
d_{82}=4.660,d_{91}=-4.350,\\
&d_{92}=8.542,
d_{101}=20.32,d_{102}=-44.68
\end{split}
\eeq

\bsp	
\label{lastpage}
\end{document}